\documentclass[preprint,12pt]{elsarticle}
\usepackage{amsthm}
\usepackage{enumerate}
\usepackage{color}
\usepackage{amsfonts}
\usepackage{tabularx}
\usepackage{mathrsfs}
\usepackage{mathdots}
\usepackage{verbatim}
\usepackage{graphicx}
\usepackage{subfigure}
\usepackage{amsmath}
\usepackage{mathabx}
\usepackage{multirow}
\usepackage{float}
\usepackage{amssymb}%

\newcounter{algorithm}
\newenvironment{algorithm}{\refstepcounter{algorithm}\vspace{1ex}
  {\sc Algorithm \thealgorithm.}\hspace{0.3em}\parindent=0pt}{\vspace{1ex}}
\newcounter{remark}

\theoremstyle{plain}
\newtheorem{theorem}{Theorem}[section]

\usepackage[%
  breaklinks=true,%
  colorlinks=true,%
  linkcolor=blue,anchorcolor=blue,%
  citecolor=blue,filecolor=blue,%
  menucolor=blue,pagecolor=blue,%
  urlcolor=blue]{hyperref}

\begin{document}

\begin{frontmatter}



\title{A New High Performance and Scalable SVD algorithm on Distributed Memory Systems}

\author[work1]{Shengguo Li\corref{cor} }
\cortext[cor]{Corresponding author} \ead{nudtlsg@nudt.edu.cn}
\author[work1]{Jie Liu}
\author[work2]{Yunfei Du}


\address[work1]{College of Computer, National University of Defense Technology, Changsha 410073, China}

\address[work2]{Department of Computer Science, Sun Yat-Sen University, Guangzhou 510006, China}





\begin{abstract}
This paper introduces a high performance implementation of \texttt{Zolo-SVD} algorithm
on distributed memory systems, which is based on the polar decomposition (PD) algorithm
via the Zolotarev's function (\texttt{Zolo-PD}),
originally proposed by Nakatsukasa and Freund [SIAM Review, 2016].
Our implementation highly relies on the routines of ScaLAPACK and therefore it is portable.
Compared with the other PD algorithms such as the QR-based dynamically weighted Halley method (\texttt{QDWH-PD}),
\texttt{Zolo-PD} is naturally parallelizable and has better scalability though performs more floating-point operations.
When using many processes, \texttt{Zolo-PD} is usually 1.20 times faster than \texttt{QDWH-PD} algorithm, and
\texttt{Zolo-SVD} can be about two times faster than the ScaLAPACK routine \texttt{\texttt{PDGESVD}}.
These numerical experiments are performed on Tianhe-2 supercomputer, one of the fastest
supercomputers in the world, and
the tested matrices include some sparse matrices from particular applications and
some randomly generated dense matrices with different dimensions.
Our \texttt{QDWH-SVD} and \texttt{Zolo-SVD} implementations are freely available at
https://github.com/shengguolsg/Zolo-SVD.
\end{abstract}

\end{frontmatter}

\section{Introduction}
\label{sec:intro}

Computing the SVD of a matrix is an important problem in scientific computing
and many applications. For example, SVD has been used for information retrieval~\cite{LSI},
principal component analysis (PCA) in Statistics~\cite{hotelling-pca}, and
signal processing~\cite{moore-pca}.
How to compute it in an efficient and scalable way has gathered much attention.

State-of-the-art SVD solvers are based on the bidiagonal reduction (BRD) strategy,
consisting of the following three stages.
First, a general matrix is reduced to an upper bidiagonal form by a sequence of
(two-sided) orthogonal transformations~\cite{Demmel-book}, and this step is called \emph{bidiagonal reduction}.
Second, the bidiagonal SVD problem is solved by any standard method such as
DC~\cite{Gu-singular}, QR~\cite{DK} or MRRR~\cite{MRRR-bidiagonal}.
Finally, the singular vectors are computed by accumulating the orthogonal
transformations from the bidiagonal reduction, which is called
\emph{back transformation}.
The reduction step is the most time-consuming phase,
and takes 75\%--99\% of the total time on homogeneous multicore
architecture~\cite{Ltaif-TOMS}.
Some recent works have been focused on accelerating the bidiagonal reduction phase,
see references \cite{New-SVD,Ltaif-TOMS,Faverge-BD}.
To be efficient, much effort is required to develop an efficient and scalable algorithm.
Every detail has to be considered including synchronization of multi-threads,
vectorization, cache issues, and so on.

There are four main types of bidiagonalization methods.
One is the classical one-stage approach, used in LAPACK and ScaLAPACK, which
directly reduces a matrix to its bidiagonal form through
sequences of orthogonal transformations (Householder or Givens) from two sides.
Another is the two-stage approach, proposed in~\cite{GL99}, 
which first reduces a matrix to its
banded form and then a banded matrix is bidiagonalized.
Another approach is obtained by means of the Lanczos algorithm, see~\cite{Golub-Kahan} and \cite{Simon-Zha-bidiagonal}.
The fourth approach is the so called \emph{one-sided bidiagonalization}, first proposed by
Ralha in~\cite{Ralha-LAA}, and stabilized by Barlow, Bosner, and Drma\v{c}~\cite{Barlow-OneSided} and
a block parallel version is proposed in~\cite{Barlow-pOnesided}.

In this work we exploit a different approach instead of accelerating the bidiagonal reduction phase.
We try out some new algorithms firstly proposed in the numerical linear algebra area.
Higham and Papadimitriou~\cite{HP-polarSVD2} introduce a new framework for
computing the SVD which is based on the polar decomposition and
combines with the eigendecomposition algorithms.
Note that any rectangular matrix $A \in \mathbb{C}^{m\times n} (m\ge n)$
has a polar decomposition (PD)
\begin{equation}
  \label{eq:polar}
  A=Q_p H,
\end{equation}
where $Q_p \in \mathbb{C}^{m\times n}$ is a (tall) matrix with orthogonal columns and
$H \in \mathbb{C}^{n\times n}$ is Hermitian positive semidefinite~\cite{Higham-book2}.
The SVD of $A$ can be obtained by further computing the eigendecomposition of
$H$, .i.e, $A=Q_p (V\Lambda V^*)=(Q_pV) \Lambda V^* := U\Lambda V^*$.

One advantage of this framework is that it can be accelerated by
many efficient PD algorithms and some
well-developed scalable eigensolvers (such as
ELPA~\cite{elpa-library}) without implementing the complicated bidiagonalization codes.
Therefore, its implementation is relatively simpler.
While, its main drawback is that it requires much more floating point operations
than the bidiagonal reduction approach, see \cite{NH-qdwheig,Nakatsukasa-simrev}
for details.
This framework is well-known in the numerical linear algebra area, but
there are little or no results on its performance on supercomputers compared with
existing parallel SVD algorithms, for example, the
algorithms in \texttt{ScaLAPACK}~\cite{Scalapack,scalapack-book}, the most famous parallel numerical linear package.

The SVD problem is reduced to an eigenvalue problem via the polar decomposition, and
therefore the recent well-developed scalable eigenvalue packages are usable.
The remaining problem is how to compute the polar decomposition in a scalable and
efficient way.
There exist many distinguish algorithms such as
the scaled Newton (SN) method~\cite{Higham-book2}, the QR-based dynamically
weighted Halley (QDWH) method~\cite{NBG-Halley}, and the more recently
proposed algorithm based on Zolotarev's function~\cite{Nakatsukasa-simrev}.

The \texttt{QDWH-SVD} algorithm which is used to compute the
SVD in~\cite{NH-qdwheig},
has been implemented on multicore architecture enhanced with multiple GPUs
\cite{qdwh-multicore}.
The results there show that \texttt{QDWH-SVD} can outperform the standard methods.
Sukkari, Ltaief, and Keyes~\cite{qdwh-dist} further represent a comprehensive performance of
\texttt{QDWH-SVD} on a large-scale distributed-memory platform,
based on the numerical library ScaLAPACK~\cite{Scalapack}.
With \texttt{QDWH-PD} as a preprocessing step and using the eigensolver ELPA~\cite{elpa-library} for computing the eigendecomposition,
the distributed parallel \texttt{QDWH-SVD} algorithm~\cite{qdwh-dist} achieves up to five-fold and
two-fold than the ScaLAPACK routine \texttt{\texttt{PDGESVD}}
on ill and well-conditioned matrices, respectively.
Note that the convergence rate of QDWH relates to the condition number of matrices.
For well-conditioned matrices it requires few iterations~\cite{NH-qdwheig,NBG-Halley}
to compute the PD.
While, the condition number is usually irrelevant to the eigenvalue problems which concern more about
whether the eigenvalues are relatively well separated.

In~\cite{Nakatsukasa-simrev}, Nakatsukasa and Freund proposed a variant of
\texttt{QDWH-PD} algorithm with higher order of convergence for the polar decomposition by using
Zolotarev's function, and call it \texttt{Zolo-PD}. It is shown that the convergence order of
\texttt{Zolo-PD} can be 17, and it usually requires one or two iterations (while QDWH
requires less than 6 iterations).
As in~\cite{Nakatsukasa-simrev},
we name the SVD algorithm based on \texttt{Zolo-PD} as \texttt{Zolo-SVD}.
Up to now, we have not seen any results of \texttt{Zolo-SVD} on high performance computers.

In this paper we mainly exploit this \texttt{Zolo-SVD} algorithm, and
introduce a high performance \texttt{Zolo-PD} implementation on
distributed-memory platform based on the state-of-art numerical library
ScaLAPACK. We discuss some ways to further improve it.
We further compare the performance of \texttt{QDWH-PD} and \texttt{Zolo-PD}. It turns out that
\texttt{Zolo-PD} requires more floating point operations, while it has better scalability
than QDWH, since \texttt{Zolo-PD} decomposes MPI processes into some relatively independent groups thus
is more loosely coupled.
\texttt{Zolo-PD} can be much faster than \texttt{QDWH-PD} when using many (MPI) processes.
Combining with ELPA, we show that \texttt{Zolo-SVD} can be much faster than ScaLAPACK
routine \texttt{\texttt{PDGESVD}} and \texttt{QDWH-SVD}.
We use some sparse matrices from University of Florida
sparse matrix collection~\cite{TimDavis-Matrix}, and some randomly constructed matrices
to test the \texttt{Zolo-SVD} algorithm.
Our \texttt{QDWH-SVD} and \texttt{Zolo-SVD} so?ftware library are freely available at
https://github.com/shengguolsg/Zolo-SVD.

\section{\texttt{QDWH-PD} and \texttt{Zolo-PD}}

The polar decomposition is an important problem in numerical linear
algebra area, and the well-known PD algorithms include
the scaled Newton (SN) method~\cite{Higham-book2}, \texttt{QDWH-PD}~\cite{NBG-Halley}, and
\texttt{Zolo-PD}~\cite{Nakatsukasa-simrev}, etc.

\subsection{\texttt{QDWH-PD}}

QDWH is a QR-based dynamically weighted Halley iterative method for computing the polar
decomposition~\cite{NBG-Halley}.
QDWH computes the polar factor $Q$ as the limit of the sequence
$X_k$ defined by
\begin{equation}
  \label{eq:dwh}
  X_{k+1} = X_k(a_k I + b_k X_k^*X_k)(I+c_k X_k^*X_k)^{-1}, \quad X_0=A/\alpha,
\end{equation}
where $\alpha$ is an upper bound of the maximum singular value of $A$,
$\alpha \ge \|A\|_2$.
In QDWH the parameters $a_k, b_k$ and $c_k$ are chosen dynamically to speed
up the convergence.
If $a_k=3, b_k=1, c_k=3$ are fixed, it gives the Halley iteration, which is cubically
convergent~\cite{Higham-book2}.

The iteration~\eqref{eq:dwh} requires explicit matrix inversion and thus it may have potential
numerical stability issue.
It is shown in~\cite{NBG-Halley} that~\eqref{eq:dwh} is mathematically equivalent
to a QR-based implementation, which is inverse-free.
The practical QDWH iteration is : $X_0=A/\alpha$,
\begin{equation}
  \label{eq:qr-based}
  X_{k+1}= \frac{b_k}{c_k}X_k + \frac{1}{\sqrt{c_k}} \left(a_k -\frac{b_k}{c_k} \right)Q_1Q_2^*,
  \end{equation}
where $\begin{bmatrix} \sqrt{c_k}X_k \\ I \end{bmatrix} = \begin{bmatrix} Q_1 \\ Q_2 \end{bmatrix}R, k\ge 0.$
The main cost lies in computing the QR factorization of an $(m+n)\times n$ matrix and a matrix
multiplication, both of which can be done in a communication-optimal manner~\cite{BDHS11,Hoemmen-sisc}.

Iteration~\eqref{eq:dwh} is also mathematically equivalent to the following form,
\begin{equation}
  \label{eq:chol-based}
  \begin{split}
    Z_k & = I + c_k X_k^*X_k, \quad L_k = \text{chol}(Z_k), \\
    X_{k+1} & = \frac{b_k}{c_k}X_k + \left( a_k - \frac{b_k}{c_k} \right)(X_k L_k^{-1})L_k^{-*},
    \end{split}
\end{equation}
where chol$(Z_k)$ denotes the Cholesky factorization of $Z_k$.
The starting point is that when the absolute value of $c_k$ is small,
matrix $Z_k$ is probably well-conditioned and computing the Cholesky factorization
is cheaper than QR factorization.
As suggested in~\cite{NH-qdwheig}, we switch from~\eqref{eq:qr-based} to~\eqref{eq:chol-based} as long as
$c_k \le 100$. In general, the QR iteration~\eqref{eq:qr-based} is usually required only
once or twice.
Our implementation of the QDWH algorithm is based on ScaLAPACK, and the main procedure
is quite similar to that one in~\cite{qdwh-dist}.

\subsection{Zolo-PD}
The QDWH parameters $a, b$, and $c$ in~\eqref{eq:dwh} are computed as the solution of the rational
max-min optimization problem in~\cite{NBG-Halley},
\begin{equation}
  \label{eq:max-min}
  \max_{a,b,c} \min_{\ell \le x \le 1} x\frac{a+bx^2}{1+cx^2},
\end{equation}
subject to the constraint $f(x)=x\frac{a+bx^2}{1+cx^2} \le 1$ on $[0, 1]$.
The parameters in the QDWH iteration can also be obtained by finding the best
type-$(3, 2)$ rational approximation to the sign function in the infinity norm, see~\cite{Nakatsukasa-simrev}.

Nakatsukasa and Freund~\cite{Nakatsukasa-simrev} extend $f(x) \in \mathcal{R}_{3,2}$ to higher order rational polynomials
$f(x) \in \mathcal{R}_{2r+1, 2r}$ for general $r\ge 1$.
The obtained optimal rational function is called $Z_{2r+1}(x;\ell)$ \emph{the type$(2r+1,2r)$ Zolotarev function}
corresponding to $\ell$, and the solution is given by
\begin{equation}
  \label{eq:min-max-sol}
  Z_{2r+1}(x; \ell) = Mx \prod_{j=1}^r \frac{x^2+c_{2j}}{x^2+c_{2j-1}}.
  \end{equation}
Here, the constant $M>0$ is uniquely determined by the condition
\[
1-Z_{2r+1}(1;\ell) = -(1-Z_{2r+1}(\ell; \ell)),
\]
and the coefficients $c_1,c_2,\ldots,c_{2r}$ are given by
\begin{equation}
  \label{eq:ci}
  c_i = \ell^2 \frac{sn^2(\frac{iK^\prime}{2r+1};\ell^\prime)}{cn^2(\frac{iK^\prime}{2r+1};\ell^\prime)}, \quad i=1,2,\ldots,2r,
\end{equation}
where $sn(u;\ell^\prime)$ and $cn(u;\ell^\prime)$ are the Jacobi elliptic functions
(see, e.g., \cite[Ch. 5]{Akhiezer-book1}).
Here $\ell^\prime = \sqrt{1-\ell^2}$ and $K^\prime =\int_0^{\pi/2} \frac{d\theta}{\sqrt{1-(\ell^\prime)^2 \sin^2\theta }}$.
It is more convenient to use the scaled Zolotarev function
\begin{equation}
  \label{eq:zolotarev}
  \hat{Z}_{2r+1}(x;\ell) := \frac{Z_{2r+1}(x;\ell)}{Z_{2r+1}(1;\ell)} = \hat{M}x\prod_{j=1}^r \frac{x^2+c_{2j}}{x^2+c_{2j-1}},
\end{equation}
where $\hat{M}=\prod_{j=1}^r \frac{1+c_{2j-1}}{1+c_{2j}}$.

Combining the iterative process used in QDWH and the scaled Zolotarev functions, we arrive at an algorithm
that computes the polar factorization by composing Zolotarev functions,
called \texttt{Zolo-PD}~\cite{Nakatsukasa-simrev}, shown in Algorithm~\ref{alg:zolo_pd}.
One question remains: how to choose the order of rational functions $r$?
The convergence order is $2r+1$, and $r$ is related to the condition number of matrix $A$.
For ill-conditioned matrices, $r$ should be large for fast convergence.
A table is given in~\cite{Nakatsukasa-simrev} to guide the choice of $r$.
For completeness, the results are shown in Table~\ref{tab:choose_r}, which shows the smallest
$k$ (i.e., the number of iterations) for which
the following condition is satisfied
\[
\hat{Z}_{(2r+1)^k}([\frac{1}{\kappa_2(A)},1]; \ell)\subseteq [1-10^{-15}, 1],
\]
for varying values of $r$ and $\ell=\frac{1}{\kappa_2(A)}$.
It is suggested in~\cite{Nakatsukasa-simrev} to choose $r$ such that it requires at most
two iterations.
This strategy may require much more computational resources in the distributed parallel environment.
We will further discuss it below in section~\ref{sec:num}.

\begin{table*}%
\centering
\caption{Required number of iterations $k$ for varying $\kappa_2(A)$ and $r$}
\label{tab:choose_r}%
\begin{tabular}{c|cccccccccccc}
$\kappa_2(A)$ & 1.001  & 1.01 & 1.1 & 1.2 & 1.5 & 2 & 10 & $10^2$ & $10^3$ & $10^5$ & $10^7$ & $10^{16}$ \\ \hline
$r=1$ & 2 & 2 & 2 & 3 & 3 & 3 & 4 & 4 & 4 & 5 & 5 & 6 \\
$r=2$ & 1 & 2 & 2 & 2 & 2 & 2 & 3 & 3 & 3 & 3 & 4 & 4 \\
$r=3$ & 1 & 1 & 2 & 2 & 2 & 2 & 2 & 2 & 3 & 3 & 3 & 3 \\
$r=4$ & 1 & 1 & 1 & 2 & 2 & 2 & 2 & 2 & 2 & 3 & 3 & 3 \\
$r=5$ & 1 & 1 & 1 & 1 & 2 & 2 & 2 & 2 & 2 & 2 & 3 & 3 \\
$r=6$ & 1 & 1 & 1 & 1 & 1 & 2 & 2 & 2 & 2 & 2 & 2 & 3 \\
$r=7$ & 1 & 1 & 1 & 1 & 1 & 1 & 2 & 2 & 2 & 2 & 2 & 3 \\
$r=8$ & 1 & 1 & 1 & 1 & 1 & 1 & 2 & 2 & 2 & 2 & 2 & 2 \\
\end{tabular}
\end{table*}

The following two theorems show that equation~\eqref{eq:zolotarev} can be written in partial fraction form,
which endow \texttt{Zolo-PD} with good parallelism.
Each QR factorization in~\eqref{eq:zolo-qr} can be done simultaneously in parallel, and
therefore the computation time is reduced significantly.
For example, when $r=8$, we can divide all processes into $8$ groups, and each group
computes one QR factorization in~\eqref{eq:zolo-qr}, instead of using all processes to
compute $8$ QR factorizations sequentially.
Compared with \texttt{QDWH-PD}, another advantage of \texttt{Zolo-PD} is that it
requires much fewer iterations, about two-thirds fewer, see
Table~\ref{tab:choose_r} and~\cite{NBG-Halley},
since QDWH requires six iterations for ill-conditioned matrices.

\begin{theorem}[\cite{Nakatsukasa-simrev}]
The function $\hat{Z}_{2r+1}(x;\ell)$ as in~\eqref{eq:zolotarev} can be expressed as
  \begin{equation}
    \label{eq:zolotarev2}
    \hat{Z}_{2r+1}(x;\ell)= \hat{M}x\left( 1+ \sum_{j=1}^r \frac{a_j}{x^2+c_{2j-1}} \right),
    \end{equation}
  where
  \begin{equation}
    \label{eq:aj}
    a_j = -\left(\prod_{k=1}^r (c_{2j-1}-c_{2k}) \right)\cdotp \left( \prod_{k=1,k\ne j}^r (c_{2j-1}-c_{2k-1}) \right).
    \end{equation}
\end{theorem}

\begin{theorem}[\cite{Nakatsukasa-simrev}]
  For the function $\hat{Z}_{2r+1}(x; \ell)$ as in~\eqref{eq:zolotarev}, and a matrix $X$ with SVD
  $X=U \text{diag}(\sigma_i) V^*$, the matrix $\hat{Z}_{2r+1}(x; \ell) :=U\text{diag}(\hat{Z}_{2r+1}(\sigma_i; \ell))V^*$
  is equal to
  \begin{equation}
    \label{eq:zolotarevX2}
    \hat{Z}_{2r+1}(X; \ell) = \hat{M} \left(X +\sum_{j=1}^r a_jX(X^*X+c_{2j-1}I)^{-1}\right).
  \end{equation}
  Moreover, $\hat{Z}_{2r+1}(X;\ell)$ can be computed in an inverse-free manner as
  \begin{equation}
    \label{eq:zolo-qr}
    \begin{cases}
      \begin{bmatrix} X \\ \sqrt{c_{2j-1}}I \end{bmatrix} & =   \begin{bmatrix} Q_{j1} \\Q_{j2} \end{bmatrix} R_j, \\
      \hat{Z}_{2r+1}(X; \ell) & =  \hat{M}\left( X+\sum_{j=1}^r \frac{a_j}{\sqrt{c_{2j-1}}}Q_{j1}Q_{j2}^* \right).
      \end{cases}
    \end{equation}

\end{theorem}

\begin{algorithm}[\texttt{Zolo-PD} for the polar decomposition]
  \label{alg:zolo_pd}
Let $\alpha$ be an upper bound of $\sigma_{\max}$ of $A$ and $\beta$ a lower bound of $\sigma_{\min}$ of $X_0$.
\begin{enumerate} \setlength{\itemsep}{0pt}
\item Compute $\alpha$ and $X_0=A/\alpha$;

\item Compute $\beta$ and let $\ell_0 = \beta$;

\item Choose $r$ based on $\kappa=\ell^{-1}$ from Table~\ref{tab:choose_r}. If $\kappa < 2$ then
  $X_1=A$ and skip to (d).

\item Compute $X_1$ and $X_2$:

  \begin{itemize}

  \item[(a)] Compute $c_i$ and $a_j$ as defined in \eqref{eq:ci} and
    \eqref{eq:aj};

  \item[(b)] Compute $X_1=\hat{Z}_{2r+1}(X; \ell)$ as in \eqref{eq:zolo-qr};

  \item[(c)] Update $\ell := \hat{M}\ell \prod_{j=1}^r(\ell+c_{2j})/(\ell^2+c_{2j-1})$ and recompute $c_i$ and $a_j$ as in step (a);

  \item[(d)] Compute $X_2$ by $\hat{M} = \prod_{j=1}^r(1+c_{2j-1})/(1+c_{2j})$ and
    \[
      \scriptsize
    \begin{cases}
      Z_{2j-1} & = X_1^*X_1 + c_{2j-1}I,  L_{2j-1} = \text{Chol}(Z_{2j-1}), \\
      X_2 & = \hat{M}(X_1 + \sum_{j=1}^r a_j(X_1L_{2j-1}^{-1})L_{2j-1}^{-*}).
    \end{cases}
    \]
    Verify that $\frac{\|X_2 - X_1\|_F}{\|X_2\|_F} \le \epsilon^{1/(2r+1)}$ holds. If not, return to Step 1 with $A \leftarrow X_2$.
  \end{itemize}

\item $Q_p = X_2$ and $H=\frac{1}{2}(Q_p^*A+(Q_p^*A)^*)$.

  \end{enumerate}

\end{algorithm}

\subsection{\texttt{Zolo-SVD}}

A framework for computing the SVD via the polar decomposition and
the eigendecomposition has been proposed in~\cite{HP-polarSVD,NH-polar}.
It assumes that the polar decomposition of $A$ is $A=Q_p H$ and the symmetric eigendecomposition
of $H$ is $H=V\Sigma V^*$, and the SVD of $A$ is obtained from $A=(Q_pV) \Sigma V^*$.

Many algorithms have been proposed for this approach, and
their differences lie in how to compute the polar decomposition and the
symmetric eigendecomposition.
In~\cite{HP-polarSVD}, it suggests using a method based on Pad\'{e} iteration for
the polar decomposition and any standard method for the symmetric eigendecomposition.
In~\cite{NH-polar}, it computes the polar decomposition by the QDWH algorithm
and the symmetric eigendecomposition by QDWH-EIG which is a spectral divide-and-conquer
algorithm based on the polar decomposition, see~\cite{NH-polar} for details.
In~\cite{qdwh-multicore}, it is shown that QDWH-EIG is not as efficient as
the symmetric eigendecomposition routines in MAGMA on GPUs.
The results in~\cite{qdwh-dist} also show that
QDWH-EIG would be slower than ELPA for ill- and well-conditioned matrices on
distributed memory systems.
Therefore, we also use ELPA to compute the eigendecomposition of $H$ in our
implementation.
The SVD algorithm in~\cite{Nakatsukasa-simrev} is based on \texttt{Zolo-PD} which is implemented
in Matlab and there are no results about its performance on supercomputers.

The framework for computing SVD used in this paper is summarized in Algorithm~\ref{alg:Zolo-SVD}.
We use \texttt{Zolo-PD} to compute the polar decomposition and use ELPA~\cite{Elpa} to compute
the symmetric eigendecomposition on distributed memory parallel computers.

One of the main differences between ELPA and the DC algorithm in ScaLAPACK is that
ELPA uses two-stage approach for tridiagonalization.
Compared with ScaLAPACK, ELPA has better scalability and can be up to two times faster
on Tianhe-2 supercomputer, see~\cite{elpa-library} for results on other supercomputers.
Unfortunately, there are few packages that have efficiently implemented the two-stage bidiagonal
reduction (BRD) algorithm.
We guess that two-stage BRD extended to distributed environment systems
can achieve similar speedups as the tridiagonal reduction (TRD),
which will be our future work.
For multicore architectures, PLASMA has provided a high-performance
BRD, which achieves up to 30-fold speedup~\cite{LLD-bidiagonal} on a 16-core
Intel Xeon machine against the LAPACK implementation in Intel MKL version 10.2.

In this work, we are concerned with computing the SVD of (nearly) square matrices.
For highly rectangular matrices, many efficient algorithms are based on
randomized techniques~\cite{Martinsson-Rev10}.
Another approach is to compute the SVD \emph{incrementally}, such as the SVD-updating algorithm
in~\cite{Zha-updating}.
An hierarchically incremental SVD approach is proposed in~\cite{HI-SVD}, which is similar
in spirit to the CAQR factorization~\cite{Hoemmen-sisc}.
It divides the matrix $A$ into many chunks along the column dimension,
$A=[A_1 | A_2 | \cdots |A_m]$.
The SVD of each $A_i$ can be computed independently by using different process groups
and then merged hierarchically together.
Similar to \texttt{Zolo-PD}, it is naturally parallelizable.

\begin{algorithm}[\texttt{Zolo-SVD}]
  \label{alg:Zolo-SVD}
Input: A general matrix $A \in \mathbb{R}^{m\times n}$ with $m\ge n$. \\
Output: the SVD $A=U\Sigma V^*$.
\begin{enumerate} \setlength{\itemsep}{0pt}
\item Compute the polar decomposition $A=Q_p H$ via \texttt{Zolo-PD}.

\item Compute the symmetric eigendecomposition $H=V\Sigma V^*$ via ELPA.

\item Compute $U=Q_p V$.
\end{enumerate}

\end{algorithm}

\section{Implementation details}

The serial \texttt{Zolo-PD} algorithm requires more flops than QDWH, see~\cite{Nakatsukasa-simrev},
but it is naturally parallelizable. We can use more MPI processes and implement \texttt{Zolo-PD} in parallel.
We use $r$ process groups to compute $r$ QR factorizations or Cholesky factorizations simultaneouly.
Therefore, the operations of each iteration cost by each process group is about the same as
QDWH.
By comparing with the number of iterations required,
the parallel version of \texttt{Zolo-PD} is supposed to be about three times faster than
QDWH, which is also confirmed by the results in section~\ref{sec:num}.

In this section, we introduce our implementation details based on the routines
in ScaLAPACK, and introduce how to exploit the sparse structure of the matrix
in~\eqref{eq:zolo-qr} when computing its QR factorization.
A structured QR algorithm is proposed in the following subsection by modifying
the routines in ScaLAPACK.
It can be up to $1.5$ times faster than the QR algorithm in ScaLAPACK.

\subsection{Fast structured QR algorithms}

In Algorithm~\ref{alg:zolo_pd}, we need to compute the QR factorization of a tall matrix
$\mathcal{M}\equiv \begin{bmatrix} X \\ I \end{bmatrix}$, where
$X \in \mathbb{R}^{m\times n}$ is a general matrix, and
$I\in \mathbb{R}^{n\times n}$ is an identity matrix, sparse.
In this subsection, we investigate a fast QR algorithms for matrix $\mathcal{M}$,
which has been mentioned in~\cite[Appendix A.1]{NH-qdwheig}.
The main idea is to take advantage of the sparsity of the bottom part of
matrix $\mathcal{M}$, i.e., the identity matrix $I$.
While, the classical QR will ignore all the zeros in matrix $I$ and treat it as a dense matrix.
By exploiting this special sparse structure,
each Householder reflector in the structured QR algorithm
has at most $m+1$ nonzero elements instead of $m+n$ in the classical one.
Therefore, it could save a lot of floating point operations by exploiting this
sparse structure.
We introduce how to exploit the sparse structure of $\mathcal{M}$
block column by block column below.
Different from the method suggested in~\cite{NH-qdwheig}, our implementation can be seen as a block version of it,
and every \texttt{NB} columns are transformed together.
Therefore, each Householder reflector in our implementation
has at most $m+$\texttt{NB} nonzero elements.

\begin{figure*}[ptbh]
\centering
\includegraphics[width=4.5in,height=2.0in]{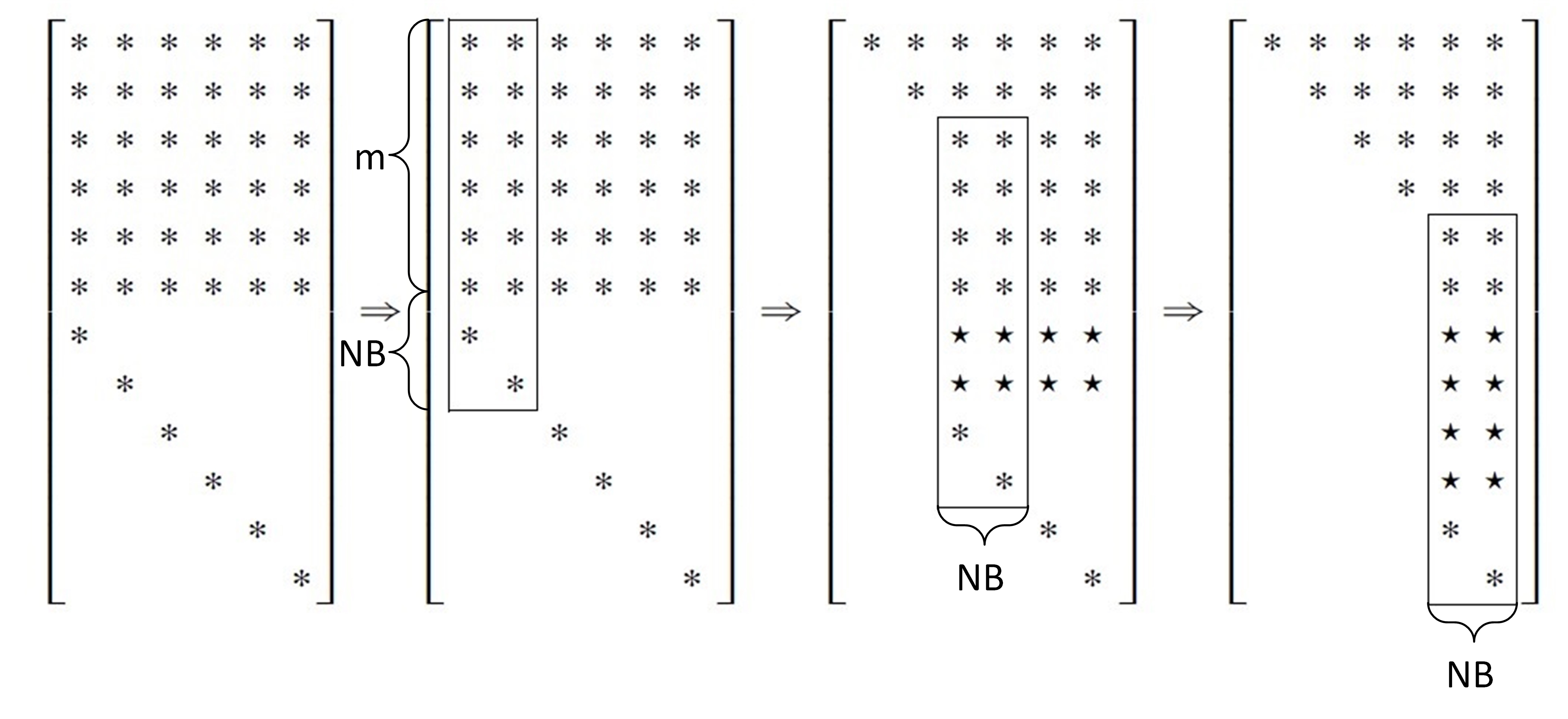}
\caption{The process of structed QR factorization}%
\label{fig:struct-qr}
\end{figure*}

In this work, we implement it by modifying the routines in ScaLAPACK, \texttt{PDGEQRF} and \texttt{PDORGQR}.
The routine \texttt{PDGEQRF} computes a QR factorization of a general matrix $A$ by
computing the QR factorization of the first \texttt{NB} columns and then the next
\texttt{NB} columns, and so on, where \texttt{NB} is the block size.
Its process is similar to that shown in Figure~\ref{fig:struct-qr} for
the proposed structured QR algorithm which is denoted by \texttt{MPDGEQRF}, for simplicity.
The difference between \texttt{MPDGEQRF} and \texttt{PDGEQRF} is that the row dimensions of the panels
in \texttt{MPDGEQRF} are at most $m$+\texttt{NB}, where \texttt{NB}$\ll n$.
Note that we use ``row dimension'' to denote the number of rows of a matrix.
While, the row dimensions of panels in \texttt{PDGEQRF} could be much larger than \texttt{MPDGEQRF}.
For example, the row dimension of the first panel of \texttt{PDGEQRF} is $m+n$.
The ScaLAPACK routine \texttt{PDORGQR} can be similarly modified to generate
the orthogonal matrix $Q$ computed by \texttt{MPDGEQRF}, and the structured version is
denoted by \texttt{MPDORGQR}.

To compare this structured QR factorization with ScaLAPACK routines,
we use two random matrices $\mathcal{M}$ with different dimensions which are
$10000\times 5000$ and $20000\times 10000$, respectively.
The experiments are done on Tianhe-2 super computer, located in Guangzhou, China.
Each compute node has two Intel Xeon E5 2692-v2 CPUs and 24 cores in total.
For the smaller matrix, we use $128$ and $256$ processes, respectively.
For the larger one, we use $512$ and $1024$ processes to test
these four routines \texttt{MPDGEQRF}, \texttt{MPDORGQR},
\texttt{PDGEQRF} and \texttt{PDORGQR}.
The execution times are shown in Table~\ref{tab:structQR}, from which we can see that
the speedups over ScaLAPACK are from $1.18$ to $1.51$.

\begin{table}%
\centering
\caption{Comparison of the structured QR algorithms with ScaLAPACK}
\label{tab:structQR}%
\begin{tabular}{c|ccc||ccc} \hline
  \multirow{2}{*}{No.Proc} & \multicolumn{6}{c}{Mat. $10000\times 5000$}  \\ \cline{2-7}
   & \texttt{PDGEQRF}  & \texttt{MPDGEQRF} & Speedup & \texttt{PDORGQR} & \texttt{MPDORGQR} &  Speedup \\ \hline
   256  & 0.91 & 0.69 & 1.32 & 0.39 & 0.27 & 1.43 \\ \hline
   512  & 0.89 & 0.66 & 1.34 & 0.30 & 0.20 & 1.51 \\ \hline
   1024 & 0.76 & 0.64 & 1.23 & 0.21 & 0.16 & 1.26 \\ \hline
   2048 & 0.87 & 0.75 & 1.18 & 0.15 & 0.13 & 1.21 \\ \hline \hline
   & \multicolumn{6}{c}{Mat. $20000\times 10000$} \\ \hline
   256  & 3.36 & 2.46 & 1.36 & 1.98 & 1.37 & 1.45 \\ \hline
   512  & 2.91 & 2.14 & 1.36 & 1.45 & 0.97 & 1.49 \\ \hline
   1024 & 2.08 & 1.60 & 1.30 & 0.88 & 0.61 & 1.46 \\ \hline
   2048 & 2.00 & 1.57 & 1.28 & 0.69 & 0.47 & 1.47 \\ \hline
\end{tabular}
\end{table}

\subsection{Implementation based on ScaLAPACK}

Our implementation highly depends on the ScaLAPACK and BLACS routines, and
therefore it is portable.
We use BLACS routines to split the communicators and perform the communications among
all these processes.
The algorithm proposed here works for both sparse and dense matrices.
Note that for sparse matrices our algorithm does not exploit their sparse properties when computing its SVD,
as done in ScaLAPACK, since Algorithm~\ref{alg:pzolosvd} can be seen as
a direct method for computing SVD.

\begin{algorithm}[Distributed Parallel \texttt{Zolo-SVD}]
  \label{alg:pzolosvd}
Input: A sparse or dense matrix $A\in \mathbb{R}^{n\times n}$.
Assume $A$ is distributed among all processes. \\
Output: The SVD of $A=U\Sigma V^T$.
\begin{enumerate} \setlength{\itemsep}{0pt}
 \item Compute $\alpha$, and $X_1=A/\alpha$, involving all the processes;
 \item Compute $\beta$ and let $\ell_0=\beta$, involving all the processes;

 \item Choose $r$ based on $\kappa=\ell^{-1}$ or fix $r=2$ or $3$.\\
 Divide all the processes into $r$ subgroups, and redistribute the
    matrix $A$ to each subgroup.

 \item While not convergent,

    \begin{itemize}
       \item[(a)] Compute $c_i$ and $a_j$ as defined in \eqref{eq:ci} and
                \eqref{eq:aj};

       \item[(b)] Compute $\hat{M} = \prod_{j=1}^r(1+c_{2j-1})/(1+c_{2j})$;

       \item[(c)] The processes in subgroup $j$ compute the QR or Cholesky factorization
       \begin{displaymath}
        \begin{bmatrix} X_1 \\ \sqrt{c_{2j-1}}I \end{bmatrix} =   \begin{bmatrix} Q_{j1} \\Q_{j2} \end{bmatrix} R_j,
        T_j=\frac{a_j}{\sqrt{c_{2j-1}}}Q_{j1}Q_{j2}^*,
        \end{displaymath}
        or
        \[
            \begin{cases}
            Z_{2j-1} & = X_1^*X_1 + c_{2j-1}I, \\
            T_j & =a_j(X_1L_{2j-1}^{-1})L_{2j-1}^{-*}), \\
            \end{cases}
         \]
       where $L_{2j-1} = \text{Chol}(Z_{2j-1}).$

       \item[(d)] The $r$ subgroups together compute
       \[
        X_2 := \hat{M}\left( X_1 + \sum_{j=1}^r T_j \right).
       \]

     \item[(e)] Verify that $\frac{\|X_2 - X_1\|_F}{\|X_2\|_F} \le \epsilon^{1/(2r+1)}$ holds. If not, $X_1 \leftarrow X_2$ and update $\ell := \hat{M}\ell \prod_{j=1}^r(\ell+c_{2j})/(\ell^2+c_{2j-1})$;
    \end{itemize}

  \item[]  {End while}

  \item One subgroup redistributes $X_2$ to all the processes;

  \item Compute the eigendecomposition of $H$, and the singular vectors of $A$.

  \end{enumerate}

\end{algorithm}

We assume the whole matrix is distributed among all the processes, and use
the BLACS routine \texttt{PDGEMR2D} to redistribute it to the process in each subcommunicator.
We find the time cost in the data redistribution is very small.
For a sparse matrix $A$, we can also let each process have one copy of $A$ when its storage is not large, stored in its sparse form.
This can avoid one time of communication.
For simplicity, we further assume an upper bound of $\sigma_{\text{max}}$ and a lower bound of
$\sigma_{\text{min}}$ are known, denoted by $\alpha$ and $\beta$, respectively.
They can be estimated in particular applications or
computed very efficiently by using some sparse direct solvers such as
SuiteSparse~\cite{TimDavis-book}, and the computation times are included in section~\ref{sec:num}, see Table~\ref{tab:sparsesuit}.
The third and fourth columns of Table~\ref{tab:sparsesuit} show the time of estimating $\alpha$ and $\beta$ in second,
which are be very small.

There are three MPI process grids in our implementation.
Assume there are $np$ processes in total and the processes are organized into a $nprow\times npcol$  2D grid, and $np=nprow\times npcol$.
The BLACS context associated with all the processes is named as \texttt{ALL\_CONTXT}.
For simpility, we assume $np$ is dividable by
$r$, $np=r\times sep\_np$.
We use the BLACS routine \texttt{BLACS\_GRIDMAP} to divide all the
processes into $r$ groups.
This is the second process grid, and its BLACS context is named as
\texttt{TOP\_CONTXT}.
Each group contains the processes in the same row of the process grid \texttt{TOP\_CONTXT}.
The communications in the \texttt{Zolo-PD} algorithm are among the processes in the same column
of the process grid \texttt{TOP\_CONTXT}.
The processes in the same row of \texttt{TOP\_CONTXT} are further organized
into a $sprow\times spcol$ 2D grid which are used to compute each
individual QR or Cholesky factorizations in~\eqref{eq:zolo-qr}.
This is the third process grid and its BLACS context is named as
\texttt{SEP\_CONTXT}.
The matrix $A$ is redistributed from the processes in \texttt{ALL\_CONTXT} to
the processes in \texttt{SEP\_CONTXT} by using the BLACS routine
\texttt{PDGEMR2D}.
The parallel \texttt{Zolo-SVD} algorithm is summarized in Algorithm~\ref{alg:pzolosvd},
which can be easily changed to an executable ScaLAPACK routine.

The main computation of \texttt{Zolo-PD} lies in the QR factorizations and the Cholesky factorizations.
Each iteration step requires a summation of $r$ terms, and the
computation of each term is independent,
which is computed by a individual subgroup of processes.
For example, it requires to compute in the QR stage
\[
X+\sum_{j=1}^r \frac{a_j}{\sqrt{c_{2j-1}}}Q_{j1}Q_{j2}^*.
\]
Each process subgroup computes the QR factorization of matrix $\mathcal{M}_i$,
\[
  \mathcal{M}_i = \begin{bmatrix} X \\ \sqrt{c_{2i-1}}I \end{bmatrix}
   =   \begin{bmatrix} Q_{i1} \\Q_{i2} \end{bmatrix} R_i.
  \]
In the Cholesky factorization stage, each process subgroup computes the Cholesky factor of
matrix $Z_{2j-1} = X_1^*X_1 + c_{2j-1}I$.
The work loads are well balanced.
To compute the summation, communications are required.
In our implementation, the summation is done by using the BLACS routine \texttt{DGSUM2D}.
To compute the summation, we only need to add the corresponding data in the same process column of
process grid \texttt{TOP\_CONTXT}.
This is because the data distribution of processes in every group is the same.

We test the performance of \texttt{Zolo-PD} by using two ways to choose the parameter $r$.
Firstly, $r$, the number of subcommunicators, is obtained from Table~\ref{tab:choose_r} based on the condition number of $A$.
This strategy works well for well-conditioned matrices.
However, for ill-conditioned matrices it requires more computational resources, for example, $r=8$.
\textbf{Another strategy} is to choose a small $r$ but use more iterations.
In our implementation, we choose $r=2$ or $3$.
From the results in Table~\ref{tab:choose_r}, the number of iterations
would increase by one or two.
This means the convergence rate of \texttt{Zolo-PD} is reduced, but its convergence rate is still of order $5$ or $7$.

\section{Numerical Results}
\label{sec:num}

Our experiments are performed on Tianhe-2 supercomputer located
in Guangzhou, China,
which is one of the fastest supercomputers in the world, having
a peak performance of 54.9 petaflops in theory and 33.86 petaflops
in Linpack benchmark.
It has a total of 16,000 compute nodes.
Each compute node is equipped with two Intel E5-2692 CPUs, 64GB of
DDR3 main memory, and
The interconnect network topology is an opto-electronic hybrid, hierarchical fat
tree.
For compilation we used Intel fortran compiler (ifort) and the optimization flag
\texttt{-O3 -mAVX},
and linked the codes to Intel MKL (composer\_xe\_2015.1.133).
As suggested in \cite{qdwh-dist}, we only investigate only pure MPI implementation,
and set NB = 64 for the two-dimensional block cyclic data distribution (BCDD).
Each compute node uses 24 MPI processes in principle.

\textbf{Example 1}
We first use three sparse matrices with medium size from real applications,
which are obtained from the {University of Florida} sparse matrix collection~\cite{TimDavis-Matrix}.
The names of these matrices are illustrated in Table~\ref{tab:sparsesuit}.
The \texttt{QDWH-PD} and \texttt{Zolo-PD} algorithms require to estimate a lower
bound of the smallest singular value $\alpha$ and an upper bound of the largest
singular value $\beta$ for these matrices.
We find that the computations for $\alpha$ and $\beta$ are quite fast,
nearly negligible. See the third and fourth columns of Table~\ref{tab:sparsesuit}, and
the experiments are performed on
a Laptop with Intel i7 CPU and 16GB memory using Matlab 2010b.

\begin{table}[pthb]
\centering
\caption{Summary of basic matrix characteristics and times of computing the lower and upper bounds}
\label{tab:sparsesuit}%
\begin{tabular}{|c|c|c|c|c|c|} \hline
  {Matrix} & {N} & $\alpha$ & $\beta$ & {Cond} & {$r$} \\ \hline
  \texttt{nemeth03}  &9,506 & 1.03e-02  & 7.19e-02  & 1.29e+00 & 2 \\
  \texttt{fv1}       & 9,604 & 3.35e-02 & 2.56e-01 & 1.40e+01 & 3 \\
  \texttt{linverse}  & 11,999 & 3.01e-03 & 0.25e-01 & 9.06e+03 & 4 \\ \hline
\end{tabular}
\end{table}

The number of iterations of \texttt{Zolo-PD} depends on $r$.
In this example, we choose $r$ from the values in Table~\ref{tab:choose_r}.
When $r$ is larger, \texttt{Zolo-PD} has higher convergence rate.
Table~\ref{tab:iteration} shows the number of iterations cost by \texttt{Zolo-PD}.
From it we can see that when $r=3 \text{ or } 4$ \texttt{Zolo-PD} requires two fewer iterations than \texttt{QDWH-PD},
which explains the speedups of \texttt{Zolo-PD} over \texttt{QDWH-PD} in some sense when implemented in parallel.


\begin{table}[ptbh]%
\caption{Time comparisons of \texttt{PDGESVD} with \texttt{Zolo-SVD}}
\label{tab:times}%
\centering
\begin{tabular}{|c|c|c|c|c|c|c|} \hline
  \multirow{2}{*}{Matrix} & \multirow{2}{*}{Method} & \multicolumn{5}{c|}{No. of Processes} \\ \cline{3-7}
    & & $256$ & $512$ & $1024$ & $2048$ & $4096$ \\ \hline
  \multirow{3}{*}{\texttt{nemeth03}} & \texttt{PDGESVD}  & 61.45 &15.97 & 15.14 & 14.85 & 15.57 \\ \cline{2-7}
                                     & \texttt{Zolo-SVD} & 24.73 & 15.11 & 10.91 & 11.83  & 10.01 \\ \cline{2-7}
                                     & Speedup  & 2.48 & 1.06 & 1.39 & 1.26 & 1.56 \\ \hline

  \multirow{3}{*}{\texttt{fv1}}      & \texttt{PDGESVD}  & 95.60 &20.18 & 19.06 & 18.11 & 19.03 \\ \cline{2-7}
                                     & \texttt{Zolo-SVD} & 30.65 & 17.58 & 12.82 & 12.65  & 11.54  \\ \cline{2-7}
                                     & Speedup  & 3.12 & 1.15 & 1.49 & 1.43 & 1.65 \\ \hline
  \multirow{3}{*}{\texttt{linverse}} & \texttt{PDGESVD}  & 144.37 & 29.87  & 26.43  & 25.01  & 32.38  \\ \cline{2-7}
                                     & \texttt{Zolo-SVD} & 50.29 & 28.64  & 19.62  & 18.64  & 15.83  \\ \cline{2-7}
                                     & Speedup  & 2.87 & 1.04 & 1.35 & 1.34 &  2.05 \\ \hline
\end{tabular}
\end{table}

The comparison results of \texttt{Zolo-SVD} and \texttt{\texttt{PDGESVD}} are shown in Table
\ref{tab:times}.
It shows that \texttt{Zolo-SVD} can be 3.12x faster than \texttt{\texttt{PDGESVD}} for matrix \texttt{fv1}
when using $256$ processes.
Note that the number of processes used by \texttt{\texttt{PDGESVD}} and
\texttt{Zolo-SVD} are shown in the first rows of Table~\ref{tab:times}.
The comparisons of \texttt{\texttt{QDWH-PD}} and \texttt{Zolo-PD} are shown in Table~\ref{tab:time2}.
It turns out that \texttt{Zolo-PD} is about 1.20x times faster than \texttt{QDWH-PD} when using
many processes.


\begin{table}[ptbh]%
\caption{The number of iterations required by \texttt{QDWH-PD} and \texttt{Zolo-PD}}
\label{tab:iteration}%
\centering
\begin{tabular}{c|c|c|c|c} \hline
  {Matrix} & QDWH  & $r=2$ & $r=3$ & $r=4$  \\ \hline
\texttt{nemeth03}  & 4  & 3 & 3 & 3  \\ \hline
 \texttt{fv1}      & 5  & 4 & 3 & 3  \\ \hline
 \texttt{linverse} & 5  & 4 & 3 & 3  \\ \hline
\end{tabular}
\end{table}

\textbf{Example 2}
The main drawback of \texttt{Zolo-PD} is that it requires too much floating-point operations for
ill-conditioned matrices when requiring at most TWO iterations.
One approach to fix this problem is to choose a small $r$.
In this example we let $r$ be $2$ and let \texttt{QDWH-PD} and \texttt{Zolo-PD} use \emph{the same number} of processes,
and the results are shown in Table~\ref{tab:time2}.
The first rows of Table~\ref{tab:time2} show the number of processes used.
From it, we can get that \texttt{Zolo-PD} is usually faster than
\texttt{QDWH-PD} when using the same number of processes.
Because \texttt{Zolo-PD} decomposes all MPI processes into $r$ relatively independent groups, it is more
loosely coupled and therefore more scalable than \texttt{QDWH-PD}.
\texttt{Zolo-PD} becomes faster than \texttt{QDWH-PD} when using many processes.

\begin{table}[ptbh]%
\caption{Times of \texttt{QDWH-PD} and \texttt{Zolo-PD} ($r=2$) when using the same number of processes.}
\label{tab:time2}%
\centering
\begin{tabular}{|c|c|c|c|c|c|c|c|c|} \hline
 \multirow{2}{*}{Matrix} & \multirow{2}{*}{Method} & \multicolumn{5}{c|}{No. of Processes} \\ \cline{3-7}
    & & $256$ & $512$ & $1024$ & $2048$ & $4096$ \\ \hline
  \multirow{3}{*}{\texttt{nemeth03}} & \texttt{QDWH-PD} & 16.51 & 12.02 & 8.14 & 7.36 & 5.55 \\ \cline{2-7}
                                     & \texttt{Zolo-PD} & 16.56 & 8.55 & 6.42 & 4.79 & 4.44 \\ \cline{2-7}
                                     & Speedup & 1.00 & 1.41 & 1.27 & 1.70 & 1.25 \\ \hline
  \multirow{3}{*}{\texttt{fv1}}      & \texttt{QDWH-PD} & 16.16 & 12.75 & 8.62 & 7.23 & 5.97 \\ \cline{2-7}
                                     & \texttt{Zolo-PD} & 22.26 & 10.87 & 7.40 & 6.02 & 5.82 \\ \cline{2-7}
                                     & Speedup & 0.73 & 1.17 & 1.16 & 1.20 & 1.03 \\ \hline
  \multirow{3}{*}{\texttt{linverse}} & \texttt{QDWH-PD} & 37.63 & 20.25 & 12.97 & 10.15 & 9.35 \\ \cline{2-7}
                                     & \texttt{Zolo-PD} & 37.40 & 19.07 & 12.43 & 8.75 & 7.92   \\ \cline{2-7}
                                     & Speedup & 1.01&  1.06 & 1.04 & 1.16 & 1.18 \\ \hline
\end{tabular}
\end{table}

Compared with \texttt{QDWH-PD}, \texttt{Zolo-PD} further requires communications among different
subcommunicators. We profile the \texttt{Zolo-PD} algorithm and Table~\ref{tab:time-stage} shows the times cost by each stage of \texttt{Zolo-PD},
where the rows \texttt{Combin.} show the communication time cost by \texttt{DGSUM2D}, which
computes the summation of $\sum_{j=1}^r T_j$ in Algorithm~\ref{alg:Zolo-SVD}.
The rows \texttt{FormX2} show the maximum time of computing $X_2$ of all subgroups.
The results were obtained for the matrix \texttt{fv1} when let $r=3$, and \texttt{Zolo-PD} took three
iterations, the first one used QR factorization and the other two used Cholesky factorization.
From it we can see that most time lies in computing the QR and Cholesky factorizations and
the communication times between different communicators are negligible.

\begin{table}[ptbh]
  \caption{Profiling the computational stages of \texttt{Zolo-PD} for matrix \texttt{fv1}.
  \texttt{Zolo-PD} requires three iterations and the times are in second.}
  \label{tab:time-stage}%
  \centering
\begin{tabular}{|c|c|c|c|c|c|c|} \hline
  {No. of Proc.} & $256$ & $512$ & $1024$ \\ \hline
  QR      & 3.42  & 3.03  & 2.07  \\
  Combin. & 6.64e-02 & 7.21e-03 & 6.26e-02  \\
  FormX2  & 2.82e-02 & 1.98e-02 & 1.47e-02  \\ \hline
  Chol.   & 1.95 & 1.38 & 0.89  \\
  Combin. & 1.74e-02 & 1.84e-02 & 3.80e-02  \\
  FormX2  & 1.96e-02 & 1.38e-02 & 1.26e-02  \\ \hline
  Chol.   & 1.89     & 1.22     & 0.75  \\
  Combin. & 1.34e-02 & 1.20e-02 & 1.03e-03  \\
  FormX2  & 1.87e-02 & 1.33e-02 & 1.38e-02  \\ \hline
\end{tabular}
\end{table}

%

\textbf{Example 3} We use some larger, more ill-conditioned matrices to further
test \texttt{Zolo-SVD} and compare it with \texttt{\texttt{PDGESVD}}.
The properties of these matrices are shown in Table~\ref{tab:mat2}, where
\texttt{rand1} and \texttt{rand2} are two matrices with random Gaussian entries.
These matrices include symmetric ones and nonsymmetric ones,
sparse and dense ones.
Some matrices are very ill-conditioned and their condition numbers
are in the order of $10^{11}$.

\begin{table}[ptbh]%
\centering
\caption{Summary of basic matrix characteristics}
\label{tab:mat2}%
\centering
\begin{tabular}{|c|c|c|c|} \hline
  {Matrix} & $n$ & $nnz$ & {Cond}  \\ \hline
  \texttt{bcsstk18}  & 11,948 & 80,519  & 3.46e+11 \\
  \texttt{c-47}      & 15,343 & 113,372 & 3.16e+08  \\
  \texttt{c-49}      & 21,132 & 89,087  & 6.02e+08  \\
  \texttt{cvxbqp1}   & 50,000 & 199,984 & 1.09e+11  \\
  \texttt{rand1}   & 10,000 & dense & 3.97e+07 \\
  \texttt{rand2}   & 30,000 & dense & 1.24e+07 \\ \hline
\end{tabular}
\end{table}

In this example we let \texttt{\texttt{PDGESVD}} and \texttt{Zolo-SVD} use
the same number of processes and let $r=2$, and the numerical results are
shown in Table~\ref{tab:times-large}, where the first row contains the
number of processes used and the other rows contains the speedups of \texttt{QDWH-SVD} and
\texttt{Zolo-SVD} over \texttt{PDGESVD}.
It turns out that \texttt{Zolo-SVD} is always faster than \texttt{\texttt{PDGESVD}} for these
matrices when using many MPI processes, for example using $4096$ processes.
It is interesting to see that \texttt{Zolo-SVD} is also faster than \texttt{\texttt{PDGESVD}}
when using $256$ processes.

About how to choose $r$, it depends on the computational resources you have and
the condition number of matrix, refer to Table~\ref{tab:choose_r}.
The number of iterations cost by \texttt{Zolo-PD} when choosing
different $r$ are illustrated in Table~\ref{tab:rate}, which are consistent with
the results estimated in Table~\ref{tab:choose_r}.
It only decreases by one iteration when $r$ is increased from $2$ to $5$, and
therefore $r=2$ or $3$ is probably a good choice.
Another reason is that taking $r$ too large can lead to numerical instability~\cite{Nakatsukasa-simrev}.

%
%
%
%
%

\begin{table}[tbh]%
\caption{The speedups of \texttt{QDWH-SVD} and \texttt{Zolo-SVD} ($r=2$) over \texttt{PDGESVD}}
\label{tab:times-large}%
\centering
\begin{tabular}{|c|c|c|c|c|c|c|} \hline
  \multirow{2}{*}{Matrix} &  \multirow{2}{*}{Method}& \multicolumn{5}{c|}{No. of Processes} \\ \cline{3-7}
   & & $256$  & $512$  & $1024$ & $2048$ & $4096$ \\ \hline
  \multirow{3}{*}{\texttt{bcsstk18}} & \texttt{QDWH-SVD}  & 1.61 & 0.59 & 0.76 & 0.83  & 1.02  \\ \cline{2-7}
                                     & \texttt{Zolo-SVD}  & 1.45 & 0.67 & 0.85 & 1.19 &  1.29 \\ \hline
  \multirow{3}{*}{\texttt{c-47}}     & \texttt{QDWH-SVD}  & 1.26 & 0.67  & 0.98 & 1.40 & 1.16  \\ \cline{2-7}
                                     & \texttt{Zolo-SVD}  & 2.33 & 0.92  & 1.07 &1.51 &  1.29  \\ \hline
  \multirow{3}{*}{\texttt{c-49}}     & \texttt{QDWH-SVD}  & 2.01 & 2.52  & 1.05  & 0.97  & 1.00  \\ \cline{2-7}
                                     & \texttt{Zolo-SVD}  & 1.28 & 3.02  & 1.08 &  1.54  & 1.73  \\ \hline
  \multirow{3}{*}{\texttt{cvxbqp1}}  & \texttt{QDWH-SVD}  & 1.20 & 1.12  & 1.26  & 1.03 & 1.07 \\ \cline{2-7}
                                     & \texttt{Zolo-SVD}  & 1.37 & 1.36  & 1.17 &  1.51 & 1.28 \\ \hline
  \multirow{3}{*}{\texttt{rand1}}    & \texttt{QDWH-SVD}  & 2.22 & 1.09  & 1.11  & 1.03 & 1.15 \\ \cline{2-7}
                                     & \texttt{Zolo-SVD}  & 3.21  & 1.10   & 1.24 & 1.70 &  1.79  \\ \hline
  \multirow{3}{*}{\texttt{rand2}}    & \texttt{QDWH-SVD}  & 2.17 & 2.84  & 1.14  & 1.02 & 1.73 \\ \cline{2-7}
                                     & \texttt{Zolo-SVD}  &  2.20 &  3.44 & 1.21  & 1.60 &  1.61 \\ \hline
\end{tabular}
\end{table}

Table~\ref{tab:times-large} also shows the performance of \texttt{QDWH-SVD}, which are
in the lines denoted by {\texttt{QDWH-SVD}}, the speedup compared with
\texttt{PDGESVD}. We can see that \texttt{Zolo-SVD} can compete with \texttt{QDWH-SVD}, and is usually
faster than both \texttt{QDWH-SVD} and \texttt{PDGESVD} for these matrices.
This behavior was verified for processes going from 256 up to 4096.

\subsection{Numerical Accuracy}

The backward error of the overall SVD is computed as
\begin{equation}
\label{eq:rres}
Res:=\frac{\|A-U\Sigma V^*\|_F}{\|A\|_2},
\end{equation}
where $U\in \mathbb{C}^{n\times n}$, $V\in \mathbb{C}^{n\times n}$ are orthogonal and
$\Sigma \in \mathbb{C}^{n\times n}$ is diagonal and its diagonals are the singular values,
$\|X\|_F$ denotes the Frobenius norm of matrix $X$ and
$\|X\|_2$ denotes the 2-norm of $X$.
We test the orthogonality of the computed singular vectors by measuring
\begin{displaymath}
OrthL:=\frac{\|I-UU^*\|_F}{n} \text{ and } OrthR:=\frac{\|I-VV^*\|_F}{n},
\end{displaymath}
where $U$ and $V$ are the left and right singular vectors
respectively, and $n$ is the dimension of matrix $A$,

\begin{table}%
\caption{The number of iterations cost by \texttt{Zolo-PD} when choosing different $r$}
\label{tab:rate}%
\centering
\begin{tabular}{|c|c|c|c|c|} \hline
 \multirow{2}{*}{Matrix} & \multicolumn{4}{c|}{$r$} \\ \cline{2-5}
 & $2$  & $3$  & $4$ & $5$ \\ \hline
{\texttt{bcsstk18}} & 4 & 4 & 3 & 3  \\ \hline
{\texttt{c-47}}     & 4  & 4  & 3 & 3  \\ \hline
{\texttt{c-49}}     & 4 & 4  & 3  & 3  \\ \hline
{\texttt{linverse}} & 4  & 3  & 3  & 3 \\ \hline
{\texttt{nemeth03}} & 3  & 3  & 3  & 3  \\ \hline
{\texttt{fv1}}      & 4  & 3  & 3  & 3  \\ \hline
\end{tabular}
\end{table}

\begin{figure*}[!t]
\centering
\subfigure[Res]{\includegraphics[width=2.6in]{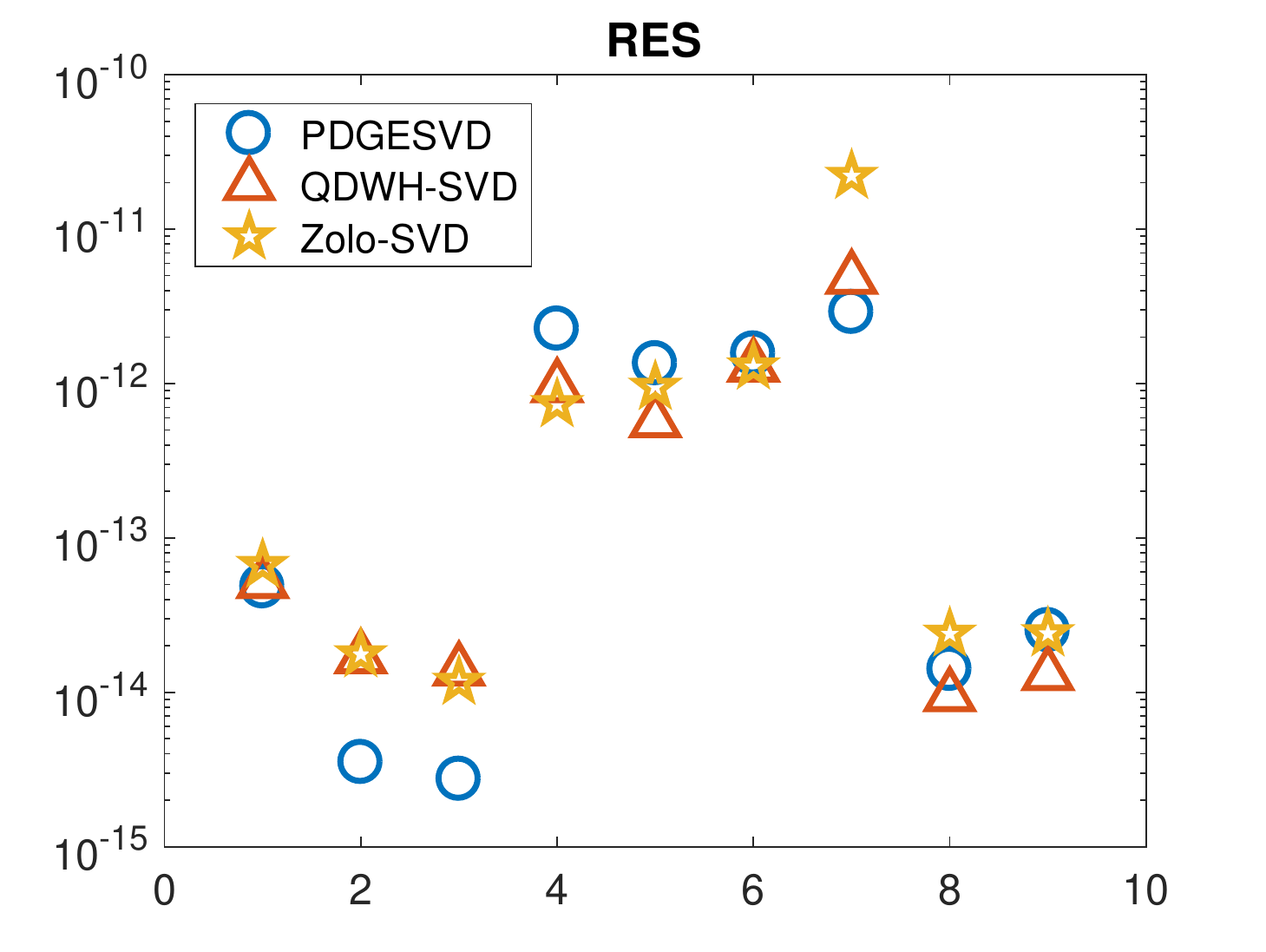}%
\label{fig:res}}
\hfil
\subfigure[OrthL]{\includegraphics[width=2.6in]{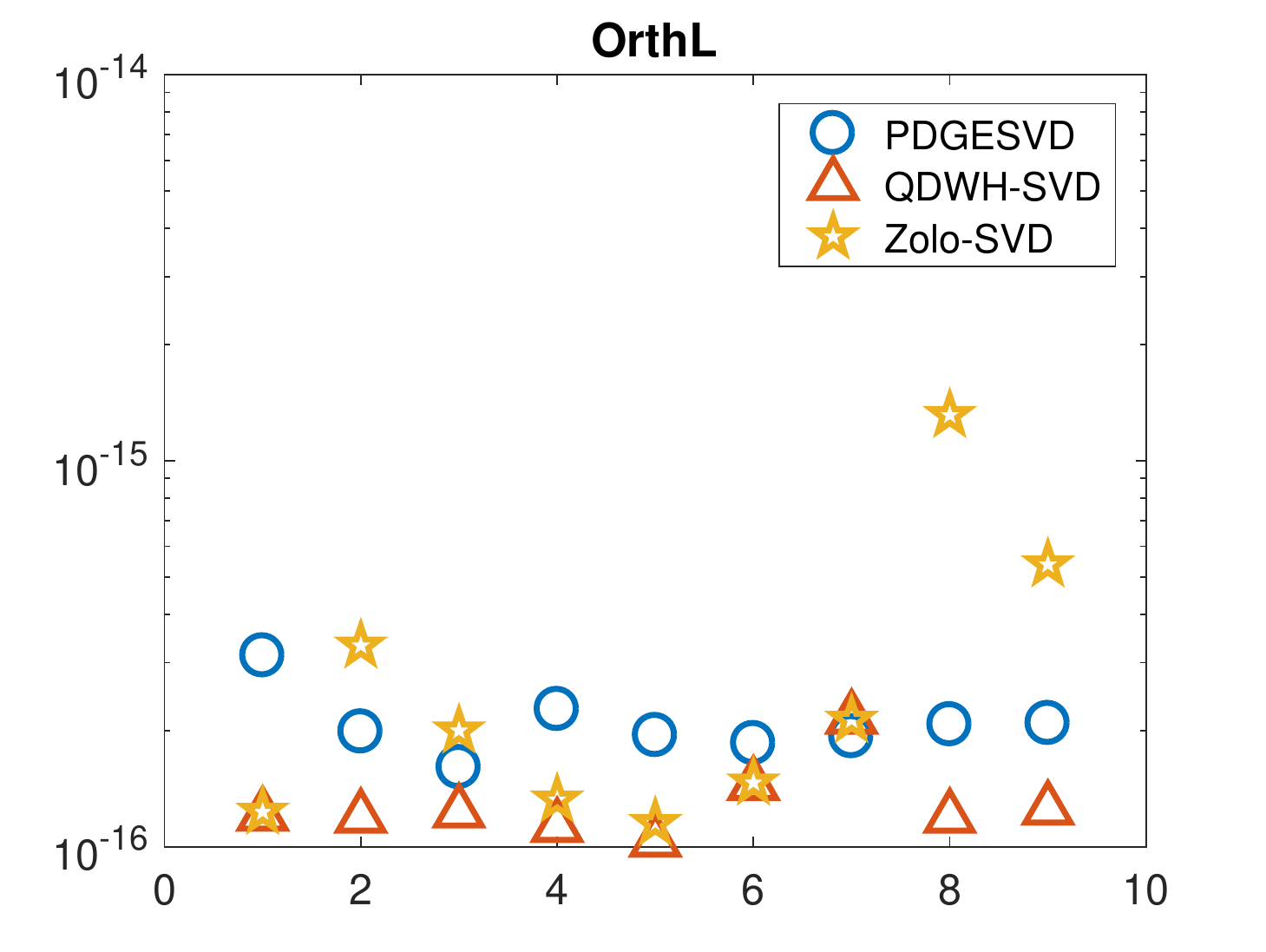}%
\label{fig:orthl}}
\caption{The residual of the computed SVD, and the orthogonality of the computed singular vectors}
\label{fig:accuracy1}
\end{figure*}

We compare these three methods, \texttt{PDGESVD}, \texttt{QDWH-SVD} and \texttt{Zolo-SVD}, and
the numerical results are shown in Figure~\ref{fig:accuracy1}.
The matrices tested are \texttt{bcsstk18},
\texttt{c-47}, \texttt{c-49}, \texttt{fv1}, \texttt{linverse}, \texttt{nemeth03},
\texttt{cvxbqp1}, \texttt{rand1}, and \texttt{rand2}, which are numbered from $1$ to $9$ in the
subfigures, respectively.
The results illustrate that the algorithm is numerically stable, and
the computed left and right singular vectors are highly orthogonal.
The relative residuals, defined in~\eqref{eq:rres}, of the computed
SVD by \texttt{Zolo-SVD} and QDWD-SVD are as accurate as those computed by \texttt{PDGESVD}
for these matrices, and the results are shown in Figure~\ref{fig:res}.
\texttt{Zolo-SVD} is comparable to \texttt{QDWH-SVD}, and their accuracy are in the same order.
The orthogonality of the computed singular vectors
by these three methods are always in machine precision,
less than $1.0e$-$15$.
The results for the right and left singular vectors are similar, and only the
results for left singular vectors are included in Figure~\ref{fig:orthl}.

\section{Conclusions}
\label{sec:conclusion}

A new distributed parallel SVD algorithm, called \emph{\texttt{Zolo-SVD}},
is implemented in this work,
which is based on the \texttt{Zolo-PD}~\cite{Nakatsukasa-simrev} algorithm
and the symmetric eigenvalue decomposition algorithm.
The main advantage of this algorithm is that it is highly scalable.
When using more computational resources, numerical results show that
\texttt{Zolo-SVD} can be three times faster than \texttt{\texttt{PDGESVD}}.
When using the same number of processes, \texttt{Zolo-SVD} can also be faster than
\texttt{\texttt{PDGESVD}}, and for some matrices it can be more than two times faster.
Compared with \texttt{QDWH-PD}, the drawback of \texttt{Zolo-PD} is that it requires much more floating point
operations. To be faster, it must be implemented in parallel.
We use many sparse matrices from the University of Florida sparse matrix collection
and some random dense matrices to conduct the numerical experiments.
Our implementations of QDWH-SVD, Zolo-SVD and structured QR algorithms are freely available at
https://github.com/shengguolsg/Zolo-SVD.

\section*{Acknowledgement}

This work is partially supported by National Natural Science Foundation of China (No. 11401580, 91530324,
91430218 and 61402495).

\end{document}